\documentclass[12pt,a4paper]{article}
\usepackage{amsmath,amscd}
\usepackage{amsfonts,amssymb}

\input{psfig.tex}

\newcommand*{\Eq}[1]{Eq.(\ref{#1})}
\author{
{\large D.Fargion},{\small $^{a}$}
\thanks{e-mail: Daniele.Fargion@roma1.infn.it}
\and{\large Yu.A.Golubkov},
{\small $^{b,c}$}\thanks{e-mail: golubkov@npi.msu.su}
\and{\large M.Yu.Khlopov},{\small $^{b,d,e}$}\thanks{e-mail: mkhlopov@orc.ru}
\and{\large R.V.Konoplich},{\small $^{b,d}$}
\thanks{e-mail: rostikon@dialup.ptt.ru}
\and{\large R.Mignani}{\small $^{f}$}
\thanks{e-mail: Roberto.Mignani@roma1.infn.it}
}

\title{{\Large{\bf Possible Effects of the Existence\\
of the 4th Generation Neutrino}}
}
\date{{\small{\it %
$^a$Dipartimento di Fisica Universit{\' a} degli Studi "La Sapienza",
00185 Rome, Italy; INFN, Sezione di Roma I\\
$^b$Center for CosmoParticle Physics ''Cosmion'',
Miusskaya sq., 4, 125047 Moscow, Russia
\\
$^c$D.V.Skobeltsyn Institute of Nuclear Physics,
M.V.Lomonosov Moscow State University,
Vorobjevy Gory, 119899 Moscow, Russia\\
$^d$Moscow State Engineering Physics Institute
(Technical University),  Kashirskoe sh., 31, 115409 Moscow, Russia\\
$^e$M.V.Keldysh Institute of Applied Mathematics,
Miusskaya sq., 4, 125047 Moscow, Russia\\
$^f$Dipartimento di Fisica "E.Amaldi" Universit{\' a} di Roma III,
00146 Rome, Italy; INFN, Sezione di Roma III
}}}

\begin{document}

\maketitle

\begin{abstract}
The 4th generation of fermions predicted by the phenomenology of heterotic
string models can possess new strictly conserved charge, which leads, in
particular, to the hypothesis of the existence of the 4th generation massive
stable neutrino. The compatibility of this hypothesis with the results of
underground experiment {\em DAMA} searching for weakly interactive particles of
dark matter and with the {\em EGRET} measurements of galactic gamma--background 
at energies above $1$ GeV fixes the possible mass of the 4th neutrino 
at the value about $50$ GeV. The possibility to test the hypothesis 
in accelerator experiments is considered. Positron signal from the annihilation 
of relic massive neutrinos in the galactic halo is calculated and is shown to be
accessible for planned cosmic ray experiments.
\end{abstract}

The superstring theory \cite{ref1} is considered in the last decade as the
approach to the finite unified ''theory of everything'', in which, in an
ideal case, all parameters of the theory are deduced from the first
principles. However there is a wide variety of possible realisations of the
superstring theory reproducing in low energy limit the Standard Model.
Therefore the analysis of superstring phenomenology plays an important role
allowing to specify possible parameters of the hidden sector of the theory.
The existence of no less than 4 fermion generations, phenomenology of broken 
$E_{6}$ symmetry (the one which includes the symmetry of the Standard
Model), $N\,=\,1$ supergravity, and (broken?) $E_{8}\,^{\prime}$ symmetry of
shadow particles and their interactions are important consequences of the
superstring theory \cite{ref2}, describing possible effects of the hidden
sector in the simplest variants of the model of heterotic string.
Identification of the 4th generation fermions with the states possessing new
strictly conserved charge leads to the stability of lightest leptons and
quarks of the 4th generation. In the present work we consider some consequences
of the hypotheses that the lightest lepton of the 4th generation is a stable
massive neutrino and we show, using in particular results of our previous
works \cite{ref3,ref4,ref5,ref6}, that the modern experimental data can be 
compatible
with this hypothesis, as well as we describe the possibilities for its
further experimental proof.

The results of measurement of $Z$ boson width exclude the possibility for the
existence of the 4th generation neutrino with a mass 
$m\,<\,m_{Z}/2\,\thickapprox\,45\ GeV$. 
Therefore the heavy neutrino mass $m$ should be greater than $m_{Z}/2$. 
Due to radiative corrections such heavy neutrinos could affect
masses of intermediate bosons. However the effect at least of one new
generation can be compensated by the increase of top quark mass within the
experimental accuracy \cite{ref7}.

In the present article we do not specify the physical nature of the 4th
generation charge but we take into account that due to conservation of this
charge the mass of heavy neutrino should be Dirac one but not Majorana mass.
Strict conservation of this charge leads to the stability of the 4th generation
neutrino.

Let us assume that the 4th generation quarks and leptons possess the same \\
$SU(3)_c\,\bigotimes\, SU(2)_L\,\bigotimes\, U(1)$ gauge charges 
as corresponding quarks and leptons of standard generations. 
In this case within the framework of
the Big Bang theory, assuming a thermodynamical equilibrium of plasma at 
$T\,>\,m$, it is possible to calculate the present number density of relic
neutrinos of the 4th generation\cite{ref4} taking into account annihilation
reactions

$$
\nu + \bar{\nu}\,\rightarrow\,f + \bar f,\ W^+ + W^-,
$$

\noindent where f denotes a light fermion.

At the stage of the formation of the Galaxy contracting 
due to its energy dissipation baryonic matter
provides an effective mechanism for the condensation of collisionless gas of
relic massive neutrinos. As a consequence an average cosmological density $%
\rho_{\nu }(0)$ of the massive neutrinos in the Galaxy increases following
the baryonic density $\rho_{b}\left( t\right) $ by the law \cite{ref3,ref4}: 

\begin{equation}
\label{eq1}
\frac{\rho_{\nu}(t)}{\rho_{\nu}(0)}
\ \sim\ \left [\frac{\rho_b(t)}{\rho_b(0)}\right ]^{\frac{3}{4}}
\end{equation}

According to \Eq{eq1} massive neutrinos are distributed mainly outside the
visible region of the Galaxy but still their central density increases up to
seven orders of magnitude with respect to the average cosmological density. Such
increase in neutrino density in the galactic halo could lead to an
observable effect of neutrino interaction with a matter in underground
detectors. As we have shown earlier \cite{ref5} the results of the
underground experiment {\em DAMA} \cite{ref8} could be compatible with 
the existence of heavy neutrino with $m\,=\,50\ GeV$. 
The range of neutrino masses 
$60\ GeV\,<\,m\,<\,290\ GeV$ is excluded by underground experiments \cite{ref5} 
if the increase of neutrino density is of the order $10^{7}$. On the other hand,
the clumping of heavy neutrinos in the galactic halo leads to an increase in
the rate of the neutrino annihilation, resulting in a copious production of
cosmic rays due to reactions 

\begin{equation}
\label{eq2}
\nu\,+\,\bar{\nu}\ \rightarrow\ \ e^+\,+\,e^-
\end{equation}

\noindent or 

\begin{equation}
\label{eq3}
\nu\,+\,\bar{\nu}\ \rightarrow\ \ q\,+\,\bar q.
\end{equation}

\begin{figure}[p] 
\centerline{%
\psfig{figure=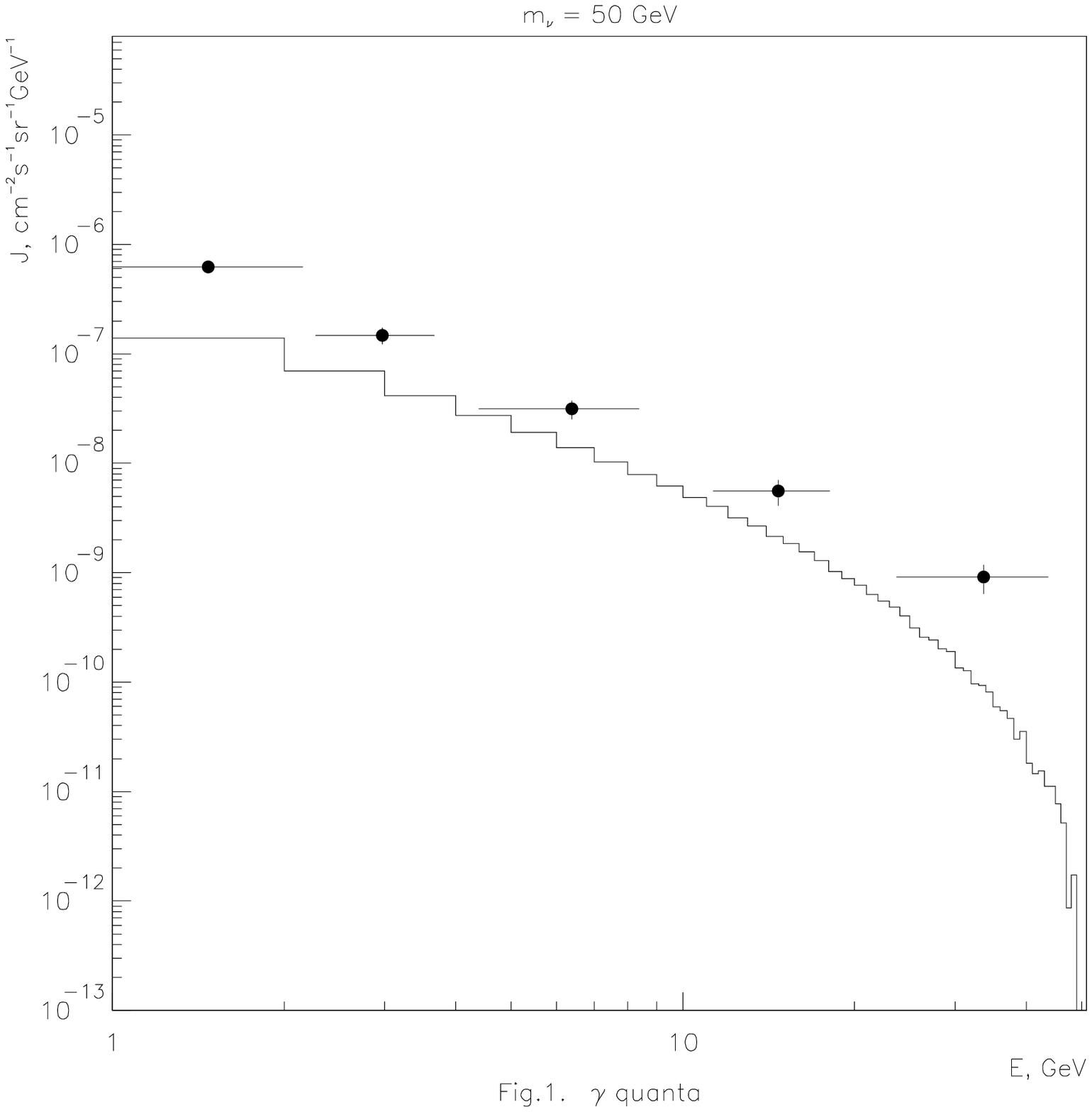,bbllx=1.0cm,bblly=5.5cm,%
bburx=18.5cm,bbury=23.5cm,clip=t,height=16.0cm}%
}%
\caption{\label{gammas}%
The comparison of the simulation of the $\gamma$--quanta flux
from the annihilation of the with $m\,=\,50\ GeV$ heavy neutrinos
in the reactions (2), (3) 
with {\em EGRET} data [9].  
The factor of the neutrino particle density enhancement in the Galaxy
in comparison with the cosmological one has been chosen
equal to $n_G/n\,=\,1.25\cdot 10^7$.
}
\end{figure}

Generally, photons are produced in hadron decays due to hadronization of
quarks and antiquarks in reaction \Eq{eq3}. To calculate the photon flux from
annihilation of heavy neutrinos in the Galaxy we used the Monte Carlo
approach described in \cite{ref6}. The results of numerical simulations of
the photon flux (see Fig.\ref{gammas}) for $m\,=\,50\ GeV$ predict 
the gamma--background of
the Galaxy below the observed by {\em EGRET} \cite{ref9} 
at $E_{\gamma }\,>\,1\ GeV$. 
However the inverse Compton effect for the energetic electrons and positrons
(produced in reaction \Eq{eq2}) on optical photons of the Galaxy \cite{ref10}
increases the photon flux by (80--40)\% in the energy range 
$1\ GeV\,<\,E_{\gamma}\,<\,15\ GeV$ \ respectively, leading to a qualitative 
agreement with {\em EGRET} data
in the energy range under consideration. Note that the {\em EGRET} observations
of gamma--background at $E_{\gamma}\,>\,50\ GeV$ indicate on the existence of
additional mechanisms for a generation of high energy gamma radiation. These
mechanisms could contribute as well at $E_{\gamma }\,<\,50\ GeV$.

High precision measurements of gamma--background in forthcoming experiments
{\hbox{{\em AGILE}}}, {\em AMS}, {\em GLAST} can give an important information 
on peculiar features of gamma-spectrum and, in particular, 
to test a sharp decrease of
gamma-spectrum at $E_{\gamma }\,\thickapprox\, 50\ GeV$ which could be 
a signature of heavy neutrino annihilation with $m\,=\,50\ GeV$.

\begin{figure}[p]
\centerline{%
\psfig{figure=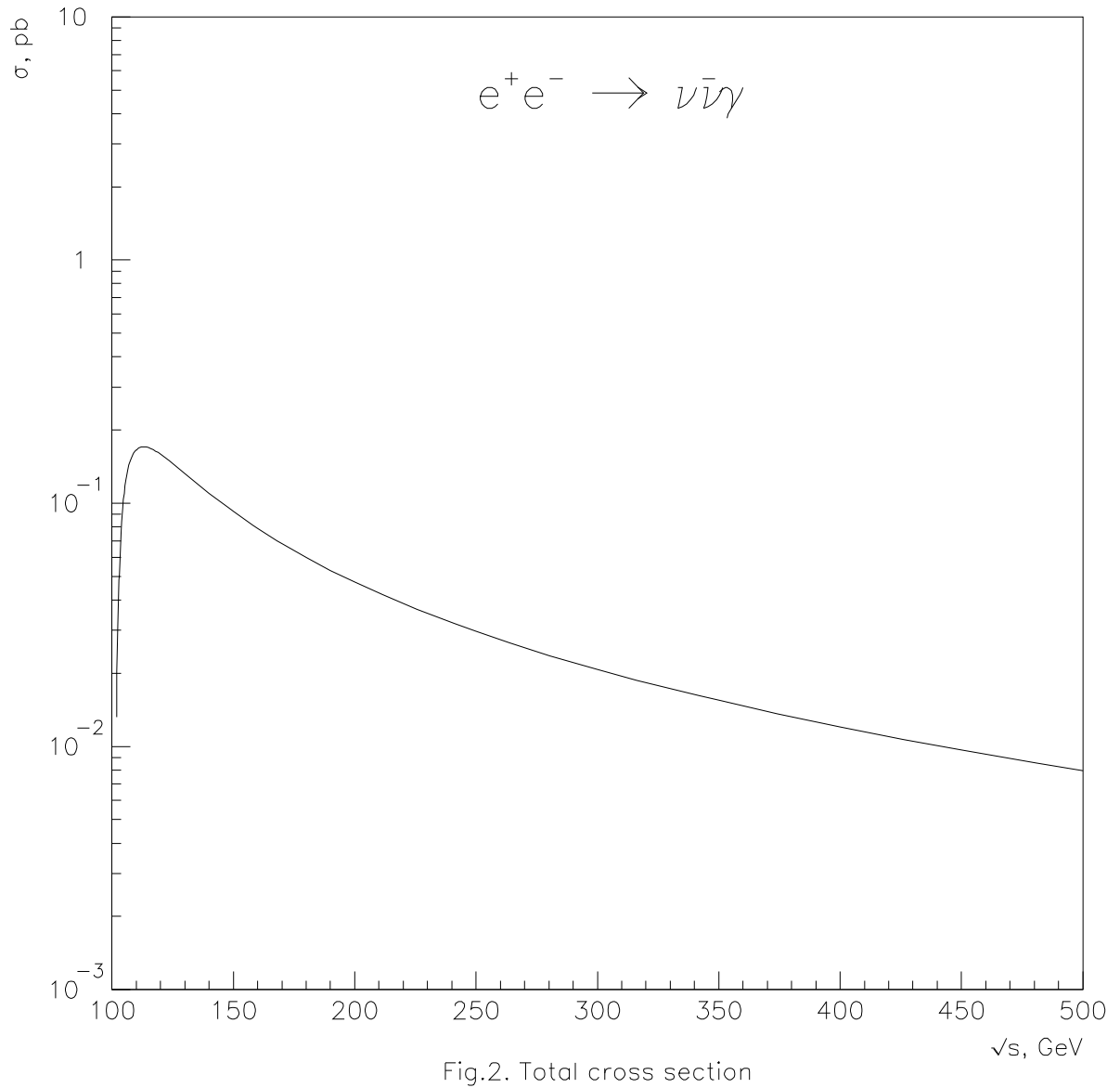,clip=t,%
bbllx=1.0cm,bblly=5.5cm,%
bburx=18.5cm,bbury=23.5cm,%
height=16.0cm}%
}%
\caption{\label{snglgam}%
The total cross section of the process
$e^+e^-\,\rightarrow\,\nu\,\bar{\nu}\,\gamma$
for the photon energy $\omega\,>\,1.5\ GeV$ and the photon angles
$30^o\,<\,\theta\,<\,150^o$.
}
\end{figure}

The hypothesis of the existence of the 4th generation neutrino with the mass
about $50\ GeV$ can be checked for example in the experiment {\em L3} on the
electron--positron collider at CERN, studying one--photon events in the
reaction $e^{+}+e^{-}\,\rightarrow\, \nu\, +\,\bar{\nu}\,+\,\gamma$ above
$Z$--resonance. In this case the 4th generation neutrinos could be observed 
due to the threshold behavior of the cross section \cite{ref11} 
(see Fig.\ref{snglgam}).

The existence of 50 GeV neutrino leads also to an interesting hadronless
signature for Higgs meson production at accelerators 

$$
e^+e^-\rightarrow\,Z\,H\rightarrow\,l^+l^-\nu\bar{\nu}
$$

\noindent and this mode would be the dominant one.

Note that if there is no new physics up to Grand Unification scale the
strong constraint on the 4th generation neutrino mass $m\,<\,220\ GeV$ 
\cite{ref7}
follows from the stability of electroweak vacuum and from the absence of
Landau pole in Higgs potential. This also stimulates the search for heavy
neutrinos in the modern accelerator experiments.

\begin{figure}[p] 
\centerline{%
\psfig{figure=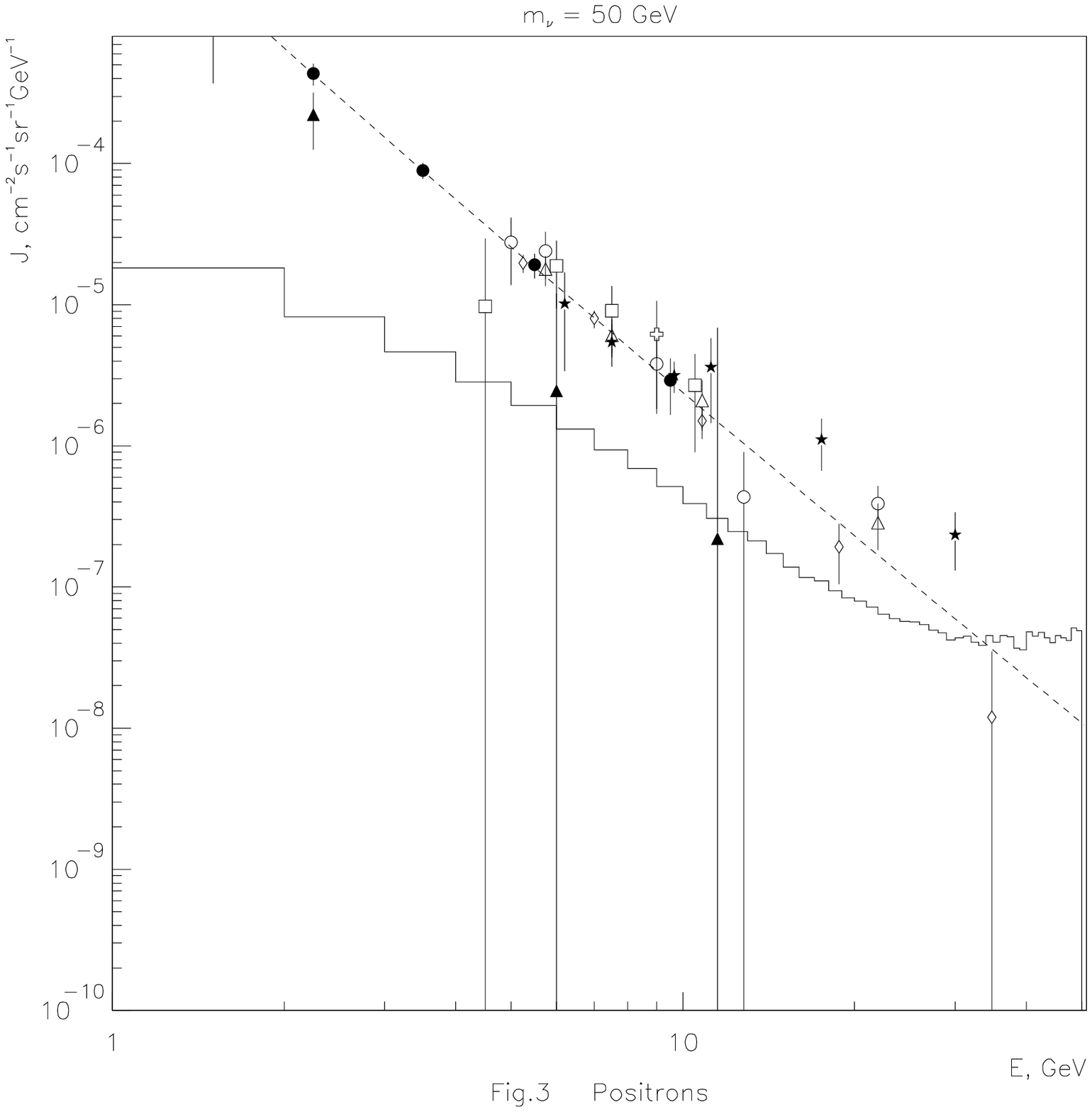,bbllx=1.0cm,bblly=5.5cm,%
bburx=18.5cm,bbury=23.5cm,clip=t,height=16.0cm}%
}%
\caption{\label{positrons}%
The simulated positron flux from the annihilation
of the $m\,=\,50\ GeV$ heavy neutrinos. 
The references on the experimental data are given in [6].
The factor of the neutrino particle density enhancement in the Galaxy
in comparison with the cosmological one has been chosen
equal to $n_G/n\,=\,1,25\cdot 10^7$.
}
\end{figure}

\begin{figure}[p]
\centerline{%
\psfig{figure=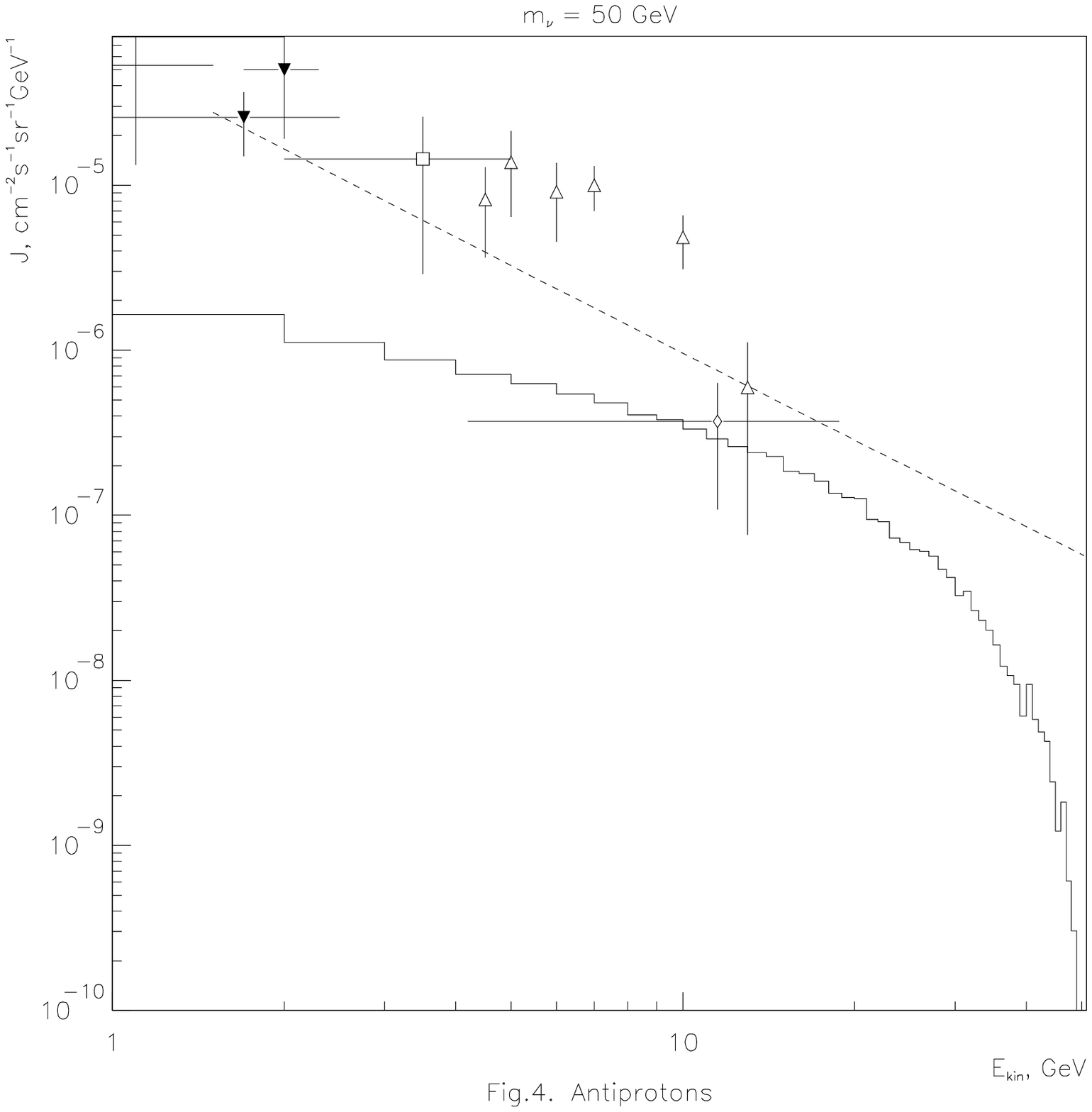,bbllx=1.0cm,bblly=5.5cm,%
bburx=18.5cm,bbury=23.5cm,clip=t,height=16.0cm}%
}
\caption{\label{aprots}%
The simulated antiproton flux from the annihilation
of the $m\,=\,50\ GeV$ heavy neutrinos. 
The references on the experimental data are given in [6].
The factor of the neutrino particle density enhancement in the Galaxy
in comparison with the cosmological one has been chosen
equal to $n_G/n\,=\,1.25\cdot 10^7$.
}
\end{figure}

Experimental measurement of positron and antiproton components of cosmic
rays, planned in the near future, in particular, in the experiment {\em AMS}
\cite{ref12}, can provide the test for the existence of primordial stable 
$50$ GeV
neutrino and antineutrino in the Galaxy by the effect of their annihilation.
The Monte Carlo simulations, described in \cite{ref6}, show 
(see Figs.\ref{positrons},\ref{aprots})
that the positron flux from the annihilation reaction \Eq{eq2} can be above
the expected flux of cosmic rays. The search for effects of the reaction 
\Eq{eq2} will allow to distinguish the physical nature of the signal (not
excluded by the results of {\em DAMA} experiment) in underground detectors. 
The interpretation of such signal by the hypothesis of a neutralino $\chi$ 
(the lightest supersymmetric particle) can not be accompanied by a significant
high energy positron signal from the annihilation of neutralinos in the
galactic halo. The reaction 

\begin{equation}
\label{eq4}
\chi\,+\,\chi\,\rightarrow\,e^+\,+\,e^-
\end{equation}

\noindent is forbidden in the s-wave due to angular momentum conservation. 
This is a
consequence of Majorana nature of the neutralino. It requires that the
annihilation should be in the p-wave, what severely suppresses the
neutralino annihilation into light fermions. Indeed, non relativistic
neutralinos in the galactic halo have relative velocities 
$v/c\,\lesssim\,10^{-3}$, and therefore the cross section of the reaction 
\Eq{eq4} (which is proportional to $(v/c)^{2}$) is suppressed.

It is important to note that the methods of testing of the hypothesis of the 4th
generation neutrino at accelerators and in experiments with cosmic rays are
complementary to each other. Indeed, if the dimensionless constant 
$\alpha_{EW}$ of heavy neutrino interaction with $Z$ bosons is suppressed 
one has to expect the corresponding suppression of heavy neutrino production at
accelerators. However the number density of relic heavy neutrino in the
Galaxy is proportional to their number density at freeze-out, which is
inversely proportional to the rate of annihilation $r\,\sim\,(\sigma v)^{-1}$.
The fluxes of particles from the annihilation in the reactions 
Eqs.(\ref{eq2}),(\ref{eq3}) are
proportional to 
$F\,\sim\, n_{\nu }n_{\bar{\nu}}\,\sim\, r^{2}(\sigma v)\,\sim\, (\sigma
v)^{-2}\,\sim\, \alpha_{EW}^{-2}$, therefore fluxes increase with the decrease
of $\alpha_{EW}$. In the opposite case the increase of the constant 
$\alpha_{EW}$ reduces the astrophysical effects of the 4th generation neutrino
but simplifies its search at accelerators.

Note that the condition for $50$ GeV neutrino lifetime to exceed the modern
age of the Universe implies very stringent conservation of the lepton charge
of the 4th generation. In particular, effective operators of dimension {\bf 5}
leading to a neutrino decay with the probability $W\,\sim\, m^{3}/m_{Pl}^{2}$
and corresponding to the life time $\tau\, <<\,t_{Univ}$, have to be excluded in
the final theory of the stable 4th generation neutrino.

Since primordial stable $50$ GeV neutrinos maintain less than $10^{-3}$ of the
critical density they can not play the dynamical role of the modern nonbaryonic 
dark matter. Thus the selfconsistent scenario of their cosmological evolution
should account for some other forms of nonbaryonic dark matter, dominating
in the modern Universe and forming the halo of our Galaxy. In the presence
of other forms of dark matter the increase of heavy neutrino density in the
Galaxy can be an order of the magnitude smaller than the one considered
here, since the effective clumping starts only after the dissipation in the
baryonic matter makes its density larger than the total density of all the
forms of dark matter. On the other hand, one can expect the increase of
neutrino density within the small scale inhomogeneities of the dominating
dark matter, what leads to the increase of their annihilation rate. It makes
the analysis of realistic scenarios for the 4th generation neutrino cosmology
the interesting subject of our further studies.

The development of the hypothesis of the 4th generation neutrino assumes the
investigation of specific properties and possible searches for the existence
of the 4th generation stable quark and stable hadrons, containing this quark. 
It would be also interesting to study the embedding of the 4th generation into
the model of horizontal unification, developed earlier \cite{ref14}. Such
possibility extends the model beyond the framework of
simplest models of heterotic string in which the rank of symmetry group does
not allow to include the gauge symmetry of generations \cite{ref15}. We can
expect that the combination of indirect experimental, cosmological and
astroparticle tests will strongly reduce the allowed set of variants for
such models.

\bigskip
This work was supported in part by the Centre for Cosmoparticle Physics
{\em ''Cosmion''} within the framework of the Section ''Cosmoparticle Physics'' 
of Russian State Scientific--Technical Programme ''Astronomy. Fundamental Space
Research'', International Project {\em ASTRODAMUS}, International collaborations
{\em EUROCOS}--{\em AMS} and {\em Cosmion}--{\em ETHZ}. Authors express 
their gratitude to
R.Battiston, V.Berezinsky, R.Bernabei, A.Dolgov, H.Hofer, and A.Starobinsky
for fruitful discussions.

\newpage


\newpage

\end{document}